%% file: main.tex
\documentclass[10pt,conference]{IEEEtran}

\usepackage{cite}
\usepackage{amsmath,amssymb,amsfonts}
\usepackage{algorithmic}
\usepackage{hyperref}
\usepackage{graphicx}
\usepackage{textcomp}
\usepackage{multirow}
\usepackage{float}
\usepackage{booktabs}
\usepackage{tabularx}

\usepackage[table,xcdraw]{xcolor}
\usepackage[normalem]{ulem}
\useunder{\uline}{\ul}{}

\newif\ifshowcomments 

\def\BibTeX{{\rm B\kern-.05em{\sc i\kern-.025em b}\kern-.08em
    T\kern-.1667em\lower.7ex\hbox{E}\kern-.125emX}}
\begin{document}

\title{A Taxonomy of Real-World Defeaters in Safety Assurance Cases\\

}

\author{\IEEEauthorblockN{Usman Gohar}
\IEEEauthorblockA{\textit{Dept. of Computer Science} \\
\textit{Iowa State University}\\
Ames, Iowa \\
ugohar@iastate.edu}
\and
\IEEEauthorblockN{Michael C. Hunter}
\IEEEauthorblockA{\textit{Dept. of Computer Science} \\
\textit{Iowa State University}\\
Ames, Iowa \\
mchunter@iastate.edu}
\and
\IEEEauthorblockN{Myra B. Cohen}
\IEEEauthorblockA{\textit{Dept. of Computer Science} \\
\textit{Iowa State University}\\
Ames, Iowa \\
mcohen@iastate.edu}
\and
\IEEEauthorblockN{Robyn R. Lutz}
\IEEEauthorblockA{\textit{Dept. of Computer Science} \\
\textit{Iowa State University}\\
Ames, Iowa \\
rlutz@iastate.edu}}

\maketitle

\input{abstract}

\input{introduction}

\input{background}
\input{methodology}
\input{taxonomy}

\input{discussion}
\input{related}
\input{conclusion}

\section*{Acknowledgment}
This work was funded by grant 80NSSC23M0058 from the National
Aeronautics and Space Administration (NASA).

{\small
\bibliographystyle{IEEEtran}
\bibliography{sample-base}}

\end{document}

%% file: abstract.tex
\begin{abstract}

The rise of cyber-physical systems in safety-critical domains calls for robust risk-evaluation frameworks. Assurance cases, often required by regulatory bodies, are a structured approach to demonstrate that a system meets its safety requirements. However, assurance cases are fraught with challenges, such as incomplete evidence and gaps in reasoning, called \emph{defeaters}, that can call into question the credibility and robustness of assurance cases. Identifying these defeaters increases confidence in the assurance case and can prevent catastrophic failures. The search for defeaters in an assurance case, however, is not structured, and there is a need to standardize defeater analysis. The software engineering community thus could benefit from having a reusable classification of real-world defeaters in software assurance cases. 
In this paper, we conducted a systematic study of literature from the past 20 years. Using open coding, we derived a taxonomy with seven broad categories, laying the groundwork for standardizing the analysis and management of defeaters in safety-critical systems. We provide our artifacts as open source for the community to use and build upon, thus establishing a common framework for understanding defeaters.
\end{abstract}

%% file: introduction.tex
\section{Introduction} 
\label{sec:introduction}

Safety-critical cyber-physical systems (CPS) pervade our lives across a variety of domains, including autonomous vehicles, robotic healthcare, and smart power grids \cite{gunes2014survey,dey2018medical,koopman2017autonomous}. These systems operate at the intersection of the digital and physical worlds, where reliability, dependability, and safety are of paramount concern \cite{hawkins2015weaving}. 
Many cyber-physical systems must undergo some form of certification or regulatory approval to ensure that they meet strict safety requirements \cite{johnson1998178b}. 
Certifying a safety-critical system typically involves constructing and submitting an \emph{assurance case} to regulators in order to demonstrate that risks are acceptable and the system will operate as intended, in compliance with applicable regulations \cite{bloomfield1998ascad}. 

\begin{figure}
    \centering
    \includegraphics[scale = 0.80]{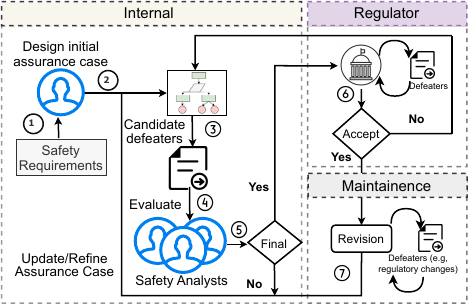}
    \caption{Overview of the safety assurance case lifecycle, 
    illustrating the roles of various stakeholders.}
    \label{fig:Defeasible}
\end{figure}

An assurance case, also called a safety assurance case, is a structured argument that the safety goal(s), i.e., the safety requirement(s), are satisfied in the delivered system operating in its intended context. For example, Knight describes an assurance case for an uncrewed aircraft, where the high-level goal (i.e., safety requirement) is that the aircraft is “adequately safe to operate in [its] prescribed environment” \cite{Knight12}. An assurance case provides a structured hierarchy of claims, arguments, and supporting evidence to show that the system will work as intended in the planned operational environment \cite{kelly2004goal,rushby2015interpretation,bloomfield2024defeaters}.

A rich body of research has introduced formal notations (e.g., Goal Structuring Notation \cite{kelly2004goal}) and various tools \cite{maksimov2019survey,10311213,denney2012advocate} to support safety analysts in constructing assurance cases. In practice, however, assurance cases are challenging to build and often suffer from incomplete, inconsistent, and unsound arguments. Such defects in assurance cases can lead to unwarranted overconfidence in system safety, with several aviation incidents linked, in part, to these shortcomings \cite{cave2006independent}.

There is thus a need to more effectively find obstacles \cite{jLamLet00} to the soundness of an assurance case's claim that its safety goals are adequately satisfied in the deployed product. These obstacles, termed \emph{defeaters}, are any factors, conditions, or events that weaken or invalidate the safety claims made about the system \cite{goodenough2015eliminative,gohar2024codefeater}. Defeaters challenge the completeness and soundness of an assurance case, often introducing uncertainty and exposing overlooked potential failures \cite{bloomfield2024defeaters}, making their identification and mitigation critical to safety assurance. 
Figure \ref{fig:Defeasible} shows a typical lifecycle of assurance cases developed from safety requirements and using defeaters. For instance, in our prior work \cite{gohar2024codefeater}, an assurance case for a small Uncrewed Aircraft System (sUAS) battery included the requirement: ``The battery charge is sufficient to complete the flight." This was challenged by the defeater: ``Unless the battery monitor is uncalibrated." Unlike fault tree analysis, which identifies events leading to a hazard, or FMECA (Failure Modes, Effects, and Criticality Analysis), which evaluates consequences of failures, defeaters specifically target potential weaknesses in assurance case arguments \cite{Knight12}.

While recent research has explored various approaches to address defeaters, such as using large language models (LLMs) with human-in-the-loop techniques \cite{gohar2024codefeater,AISupported} or semantic analysis and reasoning methods \cite{murugesan2023semantic,rushby2015understanding}, these efforts remain limited. As systems grow more autonomous and complex, anticipating all possible failure modes 
becomes increasingly difficult. Moreover, generating defeaters is labor-intensive \cite{AISupported}, relying heavily on the judgment and creativity of safety analysts, which can be prone to confirmation bias \cite{bloomfield2021safety,10311213}.  

There have been repeated calls for a systematic approach to identifying 
defeaters in assurance cases \cite{maksimov2019survey,bloomfield2021safety,rushby2015interpretation,bloomfield2024defeaters}. 
Rushby et al. \cite{rushby2015understanding} have noted that ``a systematic search for plausible defeaters may be an effective way to probe an assurance case and counteract the influence of confirmation bias." 

Towards supporting that vision, our paper proposes a taxonomy of defeaters derived from publicly available assurance cases found in the literature and our team's experience with defeater analysis. This can provide a necessary foundation for standardizing defeater analysis in assurance cases and improving the overall robustness of safety-critical systems. The proposed taxonomy can serve as a safety checklist and guide to aid developers and reviewers in improving coverage and quality of defeater analysis for safety assurance cases.

This work makes three key contributions. \textbf{1)} Surveying the real-world defeaters reported in published assurance cases;  
\textbf{2) }Using thematic analysis to derive a proposed taxonomy of defeaters from the survey’s findings; and \textbf{3) }Evaluating the defeater taxonomy in an initial application on a new real-world assurance case and suggesting potential mitigations.

The rest of the paper is organized as follows: In Section \ref{sec:background}, we provide motivation and background. Section \ref{sec:methodology} outlines our survey methodology, while Section \ref{sec:taxonomy} introduces our proposed taxonomy with examples. We discuss the implications of our work in Section \ref{sec:discussion} and related work in Section \ref{sec:related}. Section \ref{sec:conclusion} gives concluding remarks.

%% file: background.tex
\section{Motivation and Background}
\label{sec:background}
\subsection{Current practice}

\begin{figure}
    \centering
    \includegraphics[scale = 0.70]{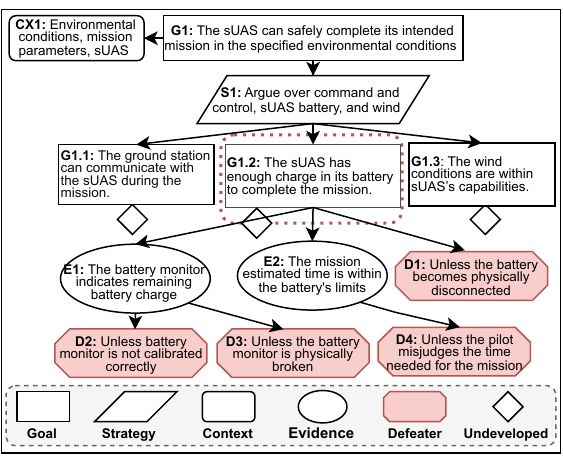}
    \caption{sUAS Safety Assurance Case fragment with example defeaters.} 
    \label{fig:AC}
    \vspace{-2.5mm}
\end{figure}

Assurance case defeaters play a critical role in certifying the safety of cyber-physical systems by identifying and addressing weaknesses in assurance cases. Defeaters represent potential doubts or objections that challenge a claim's validity, highlighting gaps in evidence and reasoning \cite{murugesan2023semantic}. Figure \ref{fig:AC} shows a fragment of an assurance case for an sUAS battery, with examples of defeaters (red boxes). Practitioners often struggle to identify defeaters systematically, as assurance cases frequently contain implicit assumptions that, if overlooked, can lead to false confidence \cite{greenwell2006taxonomy,gohar2024codefeater}. While defeater analysis strengthens the robustness of assurance cases and results in more verifiable requirements, its practical application is limited by a lack of a structured approach for identifying these defeaters. Without a systematic process, the analysis can suffer from confirmation bias and the varying expertise of reviewers \cite{maksimov2019survey,10311213}, reducing it to a superficial exercise \cite{greenwell2006taxonomy,hobbs2024driving}.

Current practices show inconsistencies; some organizations conduct a single round of defeater analysis, while others adopt an iterative process 
to target risk areas for deeper review \cite{millet2023assurance,gohar2024codefeater}. Objectives can also vary, from broad risk identification to adversarial testing or evaluating margins of error. Similar challenges exist in analogous processes (e.g., red-teaming \cite{feffer2024red}) and in various domains (e.g., AI safety), driving the development of taxonomies to better identify and manage risks \cite{ramirez2012taxonomy,mohseni2022taxonomy}.
Structured frameworks can aid practitioners’ anticipation of issues throughout the product lifecycle. As a first step towards operationalizing defeater analysis, we review the safety assurance case literature focused on creating a taxonomy of defeaters for safety-critical cyber-physical systems. The proposed taxonomy provides a systematic view of the current discourse on types of defeaters in such systems.

\begin{figure*}
    \centering
    \includegraphics[width = \textwidth]{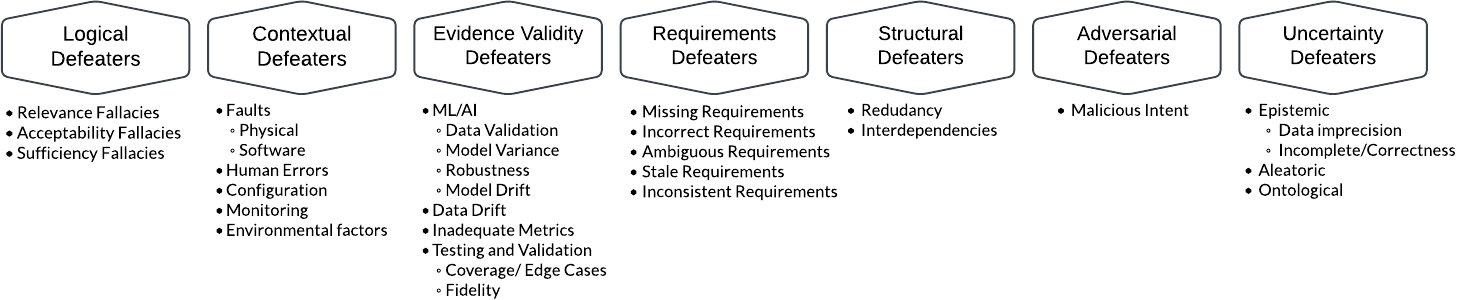}
    \vspace{-2.5mm}
    \caption{A high-level taxonomy of real-world defeaters in assurance cases. The boxes at the top show the  seven broad  categories.}
    \vspace{-2.5mm}
    \label{fig:taxonomy}
\end{figure*}

\subsection{Overview of the Automated Flight Authorization System}

Our evaluation of the proposed taxonomy is centered on an assurance case \cite{hunter2024family} we are developing for a sUAS automated flight authorization system. This system evaluates mission details, environmental conditions, and operator qualifications to determine flight permissions. It addresses the increasing operational demands and safety requirements associated with the increase in sUAS activities \cite{gohar2024towards}. The assurance case presents a series of arguments and sub-goals supporting the main claim \emph{“The sUAS can safely complete its intended mission in the specified environmental conditions.”} The system's real-world complexity makes it suitable for identifying defeaters and testing the taxonomy in practice.

%% file: methodology.tex
\section{Survey Methodology}
\label{sec:methodology}

Our approach follows prior works \cite{10.1145/3510003.3510057,shelby2023sociotechnical,gohar2024long} to review the literature and organize the taxonomy. The framework consists of three stages: (1) Identifying relevant studies; (2) Analyzing; and (3) Collating and summarizing the results into a taxonomy. 

\subsection{Identifying relevant studies}

We employed established methods \cite{greenwell2006taxonomy,shelby2023sociotechnical} to conduct a thorough search for relevant papers with publicly available safety arguments. Our initial search focused on papers published from 2004 to 2024, using the terms ``safety assurance cases'', ``safety cases'', ``defeaters'', ``fallacies'', ``doubts'' and ``assurance case weakeners'', as well as variations of these keywords, across IEEE Xplore, ACM Digital Library, and Google Scholar. We frequently encountered papers containing partial assurance cases but lacking any mention of defeaters. Therefore, we systematically reviewed each article to filter those that contained defeaters. In addition to keyword searches, we followed citation trails to and from key papers and queried Google Scholar and arXiv to identify additional related work. Finally, we incorporated safety arguments developed by our team for sUAS, resulting in six safety arguments from 10 papers. The extracted assurance case defeaters and papers are available in the accompanying artifact \footnote{\url{https://doi.org/10.5281/zenodo.14783329}}.

\subsection{Analyzing data and collating results} We used thematic analysis \cite{Clarke2014} with open-coding \cite{miles1994qualitative,gohar2024towards} to analyze the surveyed literature. Two authors independently conducted the analysis, generating new codes for each unique defeater identified. The codes were iteratively reviewed and refined to establish consensus on key themes. Overlapping codes were consolidated, and the refined taxonomy was applied to our case study. To address conceptual overlaps among codes \cite{DBLP:journals/tse/ChillaregeBCHMRW92,shelby2023sociotechnical}, the most comprehensive term was selected as the primary category covering related concepts.

\subsection{Threats to Validity}

\noindent \textbf{Sample size. } Despite our extensive survey of real-world safety arguments, the limited availability of publicly accessible safety cases constrained the sample size in this study. However, our taxonomy is designed to be flexible and extensible, allowing for the inclusion of new defeater types as they emerge. 

\noindent \textbf{Complex interdependencies. } The categories of defeaters were not always distinctly separable or orthogonal, which can result in overlaps. To mitigate this, we consolidated related defeaters with only subtle differences in their definitions to minimize redundancy in the taxonomy. 

\noindent \textbf{Evolving risks. }
Relying on existing literature may reinforce known patterns of defeaters, potentially overlooking emerging risks or novel challenges in evolving systems and technologies \cite{shelby2023sociotechnical}. As our understanding of defeaters evolves, safety experts must maintain a critical perspective and remain open to identifying new threats.

%% file: taxonomy.tex
\section{Taxonomy}
\label{sec:taxonomy}

\begin{table*}[ht]
\centering
\caption{Logical Defeaters}
\begin{tabularx}{\linewidth}{@{}l p{3.7cm} X }
\toprule
\textbf{} & \textbf{Sub-Type \cite{greenwell2006taxonomy}} & \textbf{Description and Illustrative Example} \\
\toprule
\multirow{4}{*}[-3.3ex]{\rotatebox[origin=r]{90}{Relevance}} & Appeal to Improper Authority & Referring to an authority who is not an expert in the relevant field. Example: \textit{``The sUAS is certified by NASA." NASA does not certify sUAS airframes.}\\ 
& Red Herring & Distracting the audience by introducing irrelevant content. Example: \textit{``Use the newest-looking battery." The outward appearance of the battery does not determine its health.} \\ 
& Drawing the Wrong Conclusion & The conclusion does not logically follow from the premises. Example: \textit{``The sUAS can receive signal on the ground, therefore it can receive signal in the air." This is not necessarily the case.}\\ 
& Using the Wrong Reasons & Providing irrelevant or weak reasons to support a claim. Example: \textit{``The sUAS is safe to operate because it has a high-resolution camera for accurate navigation.'' The argument cannot function as sufficient evidence of safety.}\\
\hline
\multirow{8}{*}[-9ex]{\rotatebox[origin=c]{90}{Acceptability}} & Fallacious Use of Language & Ambiguities or misleading language that affects the argument's validity. Example: \textit{``For a sUAS flight, use a battery fresh off the charger." This statement does not make clear how long the battery has been charging.}\\
& Arguing in a Circle & Using a conclusion as one of the premises. Example: \textit{``The battery won't fail if the battery works correctly." This is a restatement of the claim used as evidence.}\\ 
& Fallacy of Composition & Assuming what is true of a part is true for the whole. Example: \textit{``If the battery on an sUAS is new, the flight will be successful." There are other parts of an sUAS that could fail other than the battery. }\\ 
& Fallacy of Division & Assuming what is true of the whole is true for its parts. Example: \textit{``The sUAS previously flew successfully, so all parts function correctly." It assumes that the system's overall performance guarantees the reliability of each part.}\\ 
& False Dichotomy & Presenting only two options when more exist. Example: \textit{``Either the communications system will work or it won't." Communications might work most of the time, but not at a crucial moment.}\\ 
& Faulty Analogy & Comparing things that aren’t sufficiently alike. Example: \textit{``If a fixed-wing sUAS successfully completed a mission in this environment, a quadrotor sUAS will also." The argument doesn’t prove the systems’ differences are irrelevant.}\\ 
& Distinction without a Difference & Making an irrelevant distinction between similar things. Example: \textit{``The sUAS has been upgraded with a new flight control system." Meaningless unless arguments are supplied that demonstrate the differences between the systems.}\\ 
& Pseudo-precision & Using overly precise statistics that don't have meaningful implications. Example: \textit{``The sUAS is safe to operate as the wind forecast predicts gusts of 17.9 m/s, below the 18 m/s limit." The 
precision of .1 m/s is likely not perfect.}\\
\hline
\multirow{6}{*}[-7.5ex]{\rotatebox[origin=c]{90}{Sufficiency}} & Hasty Inductive Generalization & Drawing a conclusion based on insufficient evidence. Example: \textit{``The sUAS has sufficient power to safely complete an 8-minute flight, based on the battery consumption observed in the first two 1-minute flights.''}\\ 
& Arguing from Ignorance & Claiming something is true because it hasn't been proven false. Example: \textit{``If the battery has not had a fault before, then the battery is healthy." A lack of evidence of faults is not evidence of battery health.}\\ 
& Omission of Key Evidence & Leaving out important information that would alter the conclusion. Example: \textit{``The wind is currently less than the maximum allowed by the sUAS, therefore the wind is safe." The wind could be gusty.}\\ 
& Ignoring the Counter-Evidence & Overlooking data that contradicts the argument. Example: \textit{``The sUAS is safe because it has passed all standard safety tests.'' This ignores potential counter-evidence, like past failures in real-world conditions..}\\ 
& Confusion of Necessary and Sufficient Conditions & Misinterpreting necessary conditions as sufficient. Example: \textit{``The sUAS is safe because it has a collision avoidance system.'' While necessary, sufficient safety requires considering other factors (e.g., environmental conditions).} \\
& Gambler's Fallacy & Believing past random events affect future ones. Example: \textit{``The sUAS is safe because it has flown successfully in the last 100 flights without any issues.'' Past success does not guarantee future safety.}\\
\hline

\end{tabularx}
\label{tab:logical_defeater}
\end{table*}

Our thematic analysis identified seven broad categories of real-world defeaters. The taxonomy, presented in Figure \ref{fig:taxonomy}, focuses on common defeater categories in assurance cases for cyber-physical systems. To provide a cohesive framework for the community, we built on existing taxonomies \cite{greenwell2006taxonomy}, classifications (e.g., for uncertainty \cite{ramirez2012taxonomy}), and terminologies to avoid re-inventing new terms. This taxonomy is intended to be flexible, extensible, and evolving, with the expectation that additional defeaters, evidence, and mitigation strategies will be incorporated over time. Below, we describe each major category, highlighting subcategories and providing examples.

\subsection{Logical Fallacies}

Logical fallacies are a distinct type of defeater that can undermine reasoning in safety assurance cases, leading to flawed conclusions. These fallacies, often subtle and implicit, primarily affect the \emph{Argumentation} component of a safety case and are difficult to identify. %
Therefore, practitioners must be adept at scrutinizing argument structures to detect these underlying flaws. Greenwell \MakeLowercase{\textit{et al.}} \cite{greenwell2006taxonomy} constructed a taxonomy limited to logical fallacies, which aligns closely with our analysis of the survey and experience with assurance cases.  
To maintain consistency, we adopt their established terminology for logical-argument sub-types and integrate them into our broader taxonomy.

Greenwell \MakeLowercase{\textit{et al.}} \cite{greenwell2006taxonomy} defined three categories of logical fallacies: \emph{Relevance}, \emph{Acceptability},  and \emph{Sufficiency} fallacies. Acceptability refers to a logical flaw where the argument's premises or evidence do not have enough credibility or support to be deemed acceptable due to faulty reasoning. For instance, a \emph{circular argument} occurs when the conclusion is assumed within the premises, offering no real evidence or reasoning to substantiate the claim. Relevance fallacies occur when an argument includes information that may seem related but does not directly support the main claim, hence diverting attention and weakening the argument's logical structure. It can deceive a reviewer with irrelevant premises. Finally, sufficiency fallacies include arguments where not enough evidence is provided to support the claim. Greenwell \MakeLowercase{\textit{et al.}} \cite{greenwell2006taxonomy} further refined these three categories into the subtypes in Table \ref{tab:logical_defeater}, for each of which we provide an example for ease of understanding.

\subsection{Contextual Defeaters}

\emph{Contextual} (or operational) defeaters refer to challenges that arise when a system's predefined standards or assumptions do not hold in its operating environment. Unlike logical defeaters, which focus on logical inconsistencies, contextual defeaters impact both the evidence and claims in an assurance case by highlighting how shifts in the operational context can undermine safety arguments. Our analysis identifies several granular sub-types of contextual defeaters, shown in Table \ref{tab:contextual_defeaters} including faults (physical and software), human errors, configuration errors, monitoring failures, and environmental factors. Faults involve malfunctions in hardware, software, or subsystems due to defects, wear, or operating conditions, which can degrade system performance and impact the claim's validity. Environmental factors, such as extreme temperatures, can impact operations and must be identified to minimize risk. Human errors are a common source of defeaters and include incorrect use, bias, misinterpretation of system behavior, non-malicious operator errors, or failure to follow procedures, which can lead to safety risks. Finally, accurate monitoring is essential to ensuring safety in critical cyber-physical systems. Addressing context defeaters requires demonstrating the system's ability to detect, handle, and recover from such failures through redundancy, fault-tolerant design \cite{dubrova2013fault}, regular maintenance procedures, and adequate training.

\begin{table}[h]
\centering
\caption{Contextual Defeaters}
\begin{tabularx}{\linewidth}{@{} X }
\toprule
 \textit{\textbf{Goal:}} \textit{The sUAS battery will be adequate for the mission, considering its charge, health, ongoing monitoring, self-diagnostics, and mission duration.} \\
\end{tabularx}

\begin{tabularx}{\linewidth}{@{} p{1.9cm} X }
\toprule
\textbf{Sub-Type} & \textbf{Example \textit{(Unless...)}} \\
\midrule
Faults (Physical) & ...the battery becomes physically detached during flight. \\
Human Errors & ...the pilot miscalculates the time of the mission.\\
Configuration & ...the battery monitor is misconfigured leading to erroneous readings.\\
Monitoring & ...there is interference such that the battery level cannot be transmitted to the ground station. \\
Environmental Factors & ...the wind is stronger than anticipated, leading to rapid battery loss. \\
\bottomrule
\end{tabularx}
\label{tab:contextual_defeaters}
\end{table}

\subsection{Evidence Validity Defeaters} 

\emph{Evidence Validity} defeaters refer to challenges that compromise the trustworthiness, reliability, or completeness of the evidence used to support claims in safety assurance cases. These defeaters, shown in Table \ref{tab:evidence_integrity_defeaters}, 
undermine confidence in the data, testing results, or observations that underpin system safety evidence. They can arise from several sources, including faulty data collection, analysis errors, misinterpretation of results, or any factors that lead to incomplete, biased, or inaccurate evidence. These defeaters impact the evidence nodes of the assurance case. They can also cast doubt on the claims nodes if the evidence is insufficient or unreliable. 
With the growing integration of ML/AI in cyber-physical systems, new defeaters have emerged, including concerns about robustness, model variance, and drift, which can seriously challenge the validity of the evidence presented. Evidence validity defeaters are particularly important because they directly challenge the factual basis of the assurance case, potentially rendering the conclusions drawn from them less credible or even invalid. Possible mitigations include careful validation and verification of the evidence, especially data, robustness to perturbations, audit and review of tests, and assessing fidelity \cite{bloomfield2021safety}.

\begin{table}[t]
\centering
\caption{Evidence Validity Defeaters}
\begin{tabularx}{\linewidth}{ @{} X }
\toprule
\textit{\textbf{Goal:}} \textit{The wind conditions in the operating region (OR) are within allowed limits based on published maximum winds, forecasts, and current indicators.} \\
\end{tabularx}

\begin{tabularx}{\linewidth}{@{} l X }
\toprule
\textbf{Sub-Type} & \textbf{Example \textit{(Unless...)}} \\
\midrule
ML/AI & ...unless local wind patterns (tall buildings) were not included in the training data of the ML model.\\
Data Drift & ...the wind conditions in the OR change over time and are not detected by the pilot.\\
Inadequate Metrics & ...the metrics used to determine the sUAS's safe maximum wind speed ignore an aging airframe.\\
Testing and Validation & ...the testing used to determine maximum wind speeds did not include turbulent winds.\\
\bottomrule
\end{tabularx}
\label{tab:evidence_integrity_defeaters}
\end{table}

\subsection{Requirements Engineering Defeaters}

In assurance cases, the ``acceptably safe” goals typically assert that all safety-related requirements have been adequately met \cite{bloomfield2021safety}. This directly relies on the safety requirements being sufficient, complete, and valid to capture all necessary safety properties. Common instances of such defeaters include missing, incorrect, ambiguous, outdated (needing change), and inconsistent requirements (see Table \ref{tab:requirements_defeaters} for examples). A requirements defeater impacts the validity of the assurance case's goals, evidence, and argument structure. Possible mitigations include agile development processes to revise requirements, detailed traceability, and change-impact analysis \cite{lams09, bloomfield2021safety}.

\begin{table}[ht]
\centering
\caption{Requirements Defeaters}
\begin{tabularx}{\linewidth}{@{} X }
\toprule
\textit{\textbf{Goal:}} \textit{The ground station will control the sUAS within the transmitter's maximum range, using system diagnostics and an EM interference detector.} \\
\end{tabularx}

\begin{tabularx}{\linewidth}{@{} l X }
\toprule
\textbf{Sub-Type} & \textbf{Example \textit{(Unless...)}} \\
\midrule
Missing Reqs. & ...the diagnostics are not designed to check the communications system. \\
Incorrect Reqs. & ...the electromagnetic (EM) interference detector is not sufficiently sensitive. \\ 
Ambiguous Reqs. & ...the maximum published communications distance does not indicate if it considers tall trees or buildings 
\\
Outdated Reqs. & ...the detector cannot detect new forms of wireless communication.\\
Inconsistent Reqs. & ...the sUAS can handle less interference than the scan checks for.\\
\bottomrule
\end{tabularx}
\label{tab:requirements_defeaters}
\end{table}

\subsection{Structural Defeaters} 

\begin{table}[ht]
\centering
\caption{Structural Defeaters}
\begin{tabularx}{\linewidth}{ @{} X }
\toprule
 \textit{\textbf{Goal:}} \textit{The ground station will control the sUAS within the transmitter's maximum range, using system diagnostics and an EM interference detector.} \\
\end{tabularx}

\begin{tabularx}{\linewidth}{@{} l X }
\toprule
\textbf{Sub-Type} & \textbf{Example \textit{(Unless...)}} \\
\midrule
Lack of Redundancy & ...the transmitter is damaged by an electrical short, and there is no backup transmitter. \\
Interdependencies & ...a fault in the ground station's electric system causes multiple failures, including the transmitter.. \\
\bottomrule
\end{tabularx}
\label{tab:structural_defeaters}
\end{table}

Structural defeaters typically involve risks inherent in the system’s configuration or design, such as lack of redundancy and issues of interdependencies. These flaws or weaknesses are localized in the system’s blueprint or underlying infrastructure, which must be highlighted in the review (see Table \ref{tab:structural_defeaters}).

\subsection{Adversarial Defeaters}

Our review identified a category of defeaters caused by deliberate actions or external influences that compromise a system’s safety and reliability, which we broadly classify as \emph{Adversarial} defeaters. These defeaters can impact both the evidence and goal nodes of an assurance case, extending beyond goals specifically related to security to impact safety (see Table \ref{tab:adversarial_defeaters}). For instance, the claim that \textit{``system configurations have been verified for correct operation''} may fail to account for the risk of deliberate sabotage, exposing unrelated vulnerabilities. Prior research highlights various adversarial threats defined through security goals and requirements such as availability, confidentiality, and integrity, which can exploit weaknesses in functionality, data, or system operations \cite{HaleyLMN08}. Additionally, domain-specific taxonomies for adversarial risks such as Computer Vision \cite{long2022survey} and Machine Learning \cite{tabassi2019taxonomy} 
can be incorporated into the defeater analysis based on context. However, the detailed integration of these specialized taxonomies is beyond the scope of this work. We broadly classify these as \emph{malicious intent} defeaters to guide assurance case review.

\begin{table}[t]
\centering
\caption{Adversarial Defeaters}
\begin{tabularx}{\linewidth}{ @{} X }
\toprule
 \textit{\textbf{Goal:}} \textit{The ground station will control the sUAS within the transmitter's maximum range, using system diagnostics and an EM interference detector.} \\
\end{tabularx}

\begin{tabularx}{\linewidth}{@{} l X }
\toprule
\textbf{Sub-Type} & \textbf{Example \textit{(Unless...)}} \\
\midrule
Malicious Intent & ...the communications system is intentionally jammed by a malicious actor. \\
\bottomrule
\end{tabularx}
\label{tab:adversarial_defeaters}
\end{table}

\subsection{Uncertainty Defeaters}

Many risks in safety-critical systems arise from uncertainty, which can stem from gaps in knowledge and understanding of the system and unforeseen risks that emerge as systems evolve \cite{ramirez2012taxonomy}. \emph{Epistemic defeaters} results from gaps in knowledge or understanding of a system, which can be addressed by gathering more data or knowledge. These can be described as ``known unknowns" and include issues such as incomplete and inconsistent data. In contrast, \emph{aleatoric} defeaters refer to doubts due to inherent randomness or variability in a system, which cannot be reduced. Remedies include accounting for the intrinsic randomness by quantifying and identifying it to increase the threshold margin of error in safety-critical systems \cite{duan2017reasoning}. Finally, \emph{ontological} defeaters include factors or risks within a system that have not yet been identified or anticipated. These ``unknown unknowns" represent uncertainties about what could be present or emerge, even if they fall outside current knowledge or scope \cite{rushby2015interpretation}. (See Table \ref{tab:uncertainty_defeaters}).

\begin{table}[]
\centering
\caption{Uncertainty Defeaters}
\begin{tabularx}{\linewidth}{ @{} X }
\toprule
 \textit{\textbf{Goal:}} \textit{The SUAS’s obstacle detection system ensures safe navigation by accurately identifying and avoiding obstacles in real time.} \\
\end{tabularx}

\begin{tabularx}{\linewidth}{@{} l X }
\toprule
\textbf{Sub-Type} & \textbf{Example \textit{(Unless...)}} \\
\midrule
Epistemic & ...spurious data from the sensors leads to false negatives.\\
Aleatoric & ...random turbulence leads to failure in avoidance.  \\
Ontological & ...an unforeseen phenomenon causes the system to misidentify obstacles (unknown unknowns). \\
\bottomrule
\end{tabularx}
\label{tab:uncertainty_defeaters}
\end{table}

%% file: discussion.tex
\section{Discussion}
\label{sec:discussion}

Based on our findings, we reflect on how the taxonomy can assist stakeholders in identifying defeaters in assurance cases for cyber-physical systems, and challenges and opportunities.

\noindent \textbf{Towards structured defeater analysis.  }  
The taxonomy and its examples described here aim to provide practitioners and safety analysts with a 
structured framework for systematically identifying and understanding potential threats to an assurance case's validity. It is designed to be readily leveraged, reused, and extended. Moreover, since using different terminologies for describing similar types of defeaters undermines effective communication across different stakeholder groups, the availability of standard taxonomy categories may improve the readability of assurance cases presented to regulators. 

\noindent \textbf{Navigating tensions between known and emergent defeaters. } While the taxonomy provides an organized pathway for analyzing defeaters and reduces reliance on individual mental models, there is a risk that adherence to predefined categories may inadvertently limit practitioners' creativity, especially with novel defeaters. Anticipating emergent defeaters remains difficult, particularly in the early phases of development. In such situations, many variables remain uncertain, complicating the process. Our proposed taxonomy serves as a thematic guide that practitioners can adapt to fit specific circumstances, making it valuable for context-based analyses. Finally, the taxonomy is flexible enough to be extended as new technologies and defeaters emerge.

\noindent \textbf{Towards LLM-supported defeater analysis. } Recent studies have explored the use of Large Language Models (LLMs) for automating the detection and mitigation of defeaters in safety assurance cases \cite{gohar2024codefeater,AISupported} and related requirements tasks \cite{arora2024advancing}. While LLMs have shown promise in strengthening defeater analysis, their effectiveness is hindered by a limited scope, focusing on specific defeater types, 
and their proneness to hallucinations \cite{AISupported}. A structured taxonomy, as an external knowledge source, has been shown to be a valuable resource for enhancing the performance of LLMs 
in similar tasks \cite{zhou2024large}. Our work can potentially help LLMs to navigate complex scenarios by drawing on external knowledge, providing a more nuanced understanding of assurance-case defeaters, and suggesting mitigation strategies to improve their performance.

%% file: related.tex
\section{Related Work}
\label{sec:related}

Research efforts have developed tools to assess the structural integrity and content accuracy of assurance cases \cite{maksimov2019survey,10311213}. These tools typically focus on reviewing structural soundness (e.g., ensuring that claims lead to solutions) and providing tracking and reporting mechanisms; however, they offer limited support for comprehensive content and evidence analysis \cite{maksimov2019survey}. To evaluate the logical soundness of arguments, some studies \cite{murugesan2023semantic,muram2023attest, yuan2016automatically,rushby2015understanding} have explored predicate-based representations and semantic analysis to detect logical inconsistencies, counter-claims, and counter-evidence.
However, the majority of these works focus on logical fallacies.
More recently, some approaches \cite{gohar2024codefeater,AISupported} have sought to leverage LLMs to automatically identify and mitigate a broader range of defeaters in assurance cases. Although promising, they have limited coverage and rely on human oversight and judgment.

Several other works have sought to improve assurance case assessments, outside the context of defeaters. Yuan and Kelly \cite{yuan2012argument} introduced a framework using informal logic reasoning schemes with questions to assess the plausibility of the arguments. Luo \MakeLowercase{\textit{et al.}} \cite{luo2017systematic} proposed a general iterative process to streamline the ad-hoc nature of reviews. Separately, Chowdhury \MakeLowercase{\textit{et al.}} \cite{chowdhury2020systematic} introduced a set of criteria to evaluate the structural and content quality of an assurance case. Finally, Shahandashti \MakeLowercase{\textit{et al.}} \cite{shahandashti2024prisma} conducted a study to unify the literature on assurance case weakeners. 

%% file: conclusion.tex
\section{Conclusion}
\label{sec:conclusion}

Identifying and mitigating defeaters in software assurance cases is increasingly being adopted to increase the robustness of, and confidence in, assurance cases in safety-critical cyber-physical systems. Potential obstacles to an assurance case's claim that its safety requirements are satisfied may cast doubt on the validity and trustworthiness of the case. 
There have been growing calls to develop a framework to standardize defeater analysis. 
In this work, we argue that to facilitate this in research and practice, it is essential to develop a consolidated taxonomy of real-world defeaters. Through a review and 
thematic analysis of publicly available safety assurance cases, we provide and evaluate a proposed taxonomy of defeaters as an initial guide. 
Validation of the taxonomy with user studies is left to future work.
Our work offers a necessary step for standardizing defeater analysis in assurance cases toward improving the overall robustness of safety-critical systems. We expect and hope that the defeater taxonomy presented here will evolve as research and community engagement progress.